\documentclass[a4,12pt,epsf]{article}
\usepackage{amsmath}
\usepackage{graphicx}
\begin{document}
%%%%% title %%%%%
\begin{center}{\Large \bf
Anisotropy of dark matter velocity distribution}
\end{center}
%%%%% author(s) %%%%%
\begin{center}
Keiko I. Nagao%$^a$
\footnote{nagao@dap.ous.ac.jp}
\footnote{This report is based on the paper \cite{Nagao:2017yil}.} 
\vspace{6pt}\\
%%%%% address(es) %%%%%
%$^a$
{\it
Okayama University of Science,
1-1 Ridaicho, Kita-ku, Okayama-shi 700-0005 Japan
}
% Anoterh Address
%{
%\\
%$^b$
%{\it  another address
%Yukawa Institute for Theoretical Physics, YITP,
%Kyoto 606-8502, Japan
%}
\end{center}
%%%%% abstruct %%%%%
\begin{abstract}
Direct detection of dark matter with directional sensitivity has the potential to discriminate the dark matter velocity distribution.
Especially, it will be suitable to discriminate isotropic distribution from anisotropic one.
Analyzing data produced with Monte-Carlo simulation, required conditions for the discrimination is estimated. 
If energy threshold of detector is optimized, $O(10^3-10^4)$ event number is required to discriminate the anisotropy.
\end{abstract}
%%%%%%%%%%%%%%%%%%%%
\section{Introduction}
The so-called dark matter accounts for about $27$\% of the energy density of the Universe.
Since it cannot be directly observed, it is supposed to have exceedingly weak interaction with 
the standard model particles. Weak interacting massive particles (WIMPs) are a promising candidate
for dark matter. Several experiments are optimized to direct search for the WIMPs. Directional direct detections
of dark matter aims to detect both the recoil energy and direction of the nuclear recoils. The directional detection
is expected to improve the background rejection efficiency, and furthermore, to obtain other information of dark matter
such as the velocity distribution. 

In most of direct searches, the velocity distribution of dark matter is supposed to be isotropic Maxwellian velocity distribution. 
However, non-Maxwellian distribution had been indicated by some simulations and observations [2-7]. 
In this study, an anisotropic velocity distribution derived in \cite{LNAT} is adopted;
\begin{align}
f(v_\phi) = \frac{1-r}{N(v_{0,\mathrm{iso.}})} \exp\left[-v_\phi^2/v_{0,\mathrm{iso.}}^2\right] + 
		\frac{r}{N(v_{0,\mathrm{ani.}})} \exp\left[-(v_\phi-\mu)^2/v_{0,\mathrm{ani.}}^2\right],
\end{align}
where $v_\phi$ is the tangential velocity of dark matter with respect to the galactic rest frame, $r$ is a parameter associated with the anisotropy, $N(v)$ is normalization factor, and $\mu=150$ km/s. The radial velocity $v_r$ and velocity across the galactic plane $v_z$  are
suggested to be the isotropic Maxwellian distribution. 

%The velocity distribution of dark matter, especially its anisotropy, can be in principle measured with the directional dark matter search. 
This paper is organized as follows: in Section \ref{sec:numericalsimulation}, setup of simulation of dark matter-nucleon scattering in directional detector
is described. Results of the numerical simulation are also presented in the section. We conclude in Section \ref{sec:conclusion}.

\section{Numerical simulation}
\label{sec:numericalsimulation}
\begin{figure}[htbp] %  figure placement: here, top, bottom, or page
   \centering
   \includegraphics[width=2in]{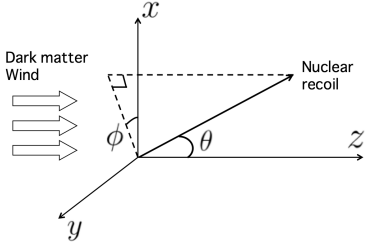} 
   \caption{Direction of a nuclear recoil}
   \label{fig:labsys}
\end{figure}
%Both the recoil energy and direction of nuclear recoil are obtained in directional detector. 
In Figure \ref{fig:labsys}, 
a nuclear recoil and associated angles in laboratory frame are shown. The Earth receives dark matter wind, and its direction is 
taken as z-axis. Scattering angle $\theta$ is defined as the angle from z-axis. As a result of Monte-Carlo simulation of dark matter-nucleon 
scatterings, both the recoil energy $E_R$ and the scattering angle  $\theta$ are obtained. Thus, in principle, the energy-angular distribution is available. It is also interesting to take a look at angular histogram, which can be obtained in the case that energy resolution of the detector is not so good. 

As a target nucleon, two typical target, fluorine (F) and silver (Ag), are supposed in the simulation. Fluorine is used in gaseous directional detectors, and silver is one of target nucleons in NEWSdm. The strategy is as follows: two kinds of dataset are generated in the Monte-Carlo simulation. 
One dataset has a large event number, and called as ``template data''. 
Template data is produced depending on the anisotropy parameter $r$, like $r=0$. $r=0.1$, $r=0.2$, $\cdots$, $r=1$.
The other dataset, which is called as ``pseudo-experimental data'', is
supposed to be data obtained in the realistic experiment, and has smaller event number than template data. 
Questions are which template is most similar to pseudo-experimental data, and how much event number is required to estimate it. 
Energy-angular distributions for template data and pseudo-experimental data are produced, and their similarity is tested by chi-squared test 
in Subsection \ref{subsec:distribution}. In the chi-squared test, 
$E_R$-$\cos{\theta}$ plane is divided into small bins, and event numbers in each bin.
Corresponding chi-squared test for angular histograms is shown in Subsection \ref{subsec:histogram}. 
For both the energy-angular distribution and angular histogram, 
mass relation between WIMP mass $m_\chi$ and target nucleon $m_N$ is supposed to be $m_\chi=3m_N$ for similicity.
Also energy threshold of the detector is supposed to be $20$ keV for target F, and $50$ keV for target Ag.

\subsection{Energy-angular distribution}
\label{subsec:distribution}
\begin{figure}[htbp] %  figure placement: here, top, bottom, or page
   \centering
   \includegraphics[width=15cm]{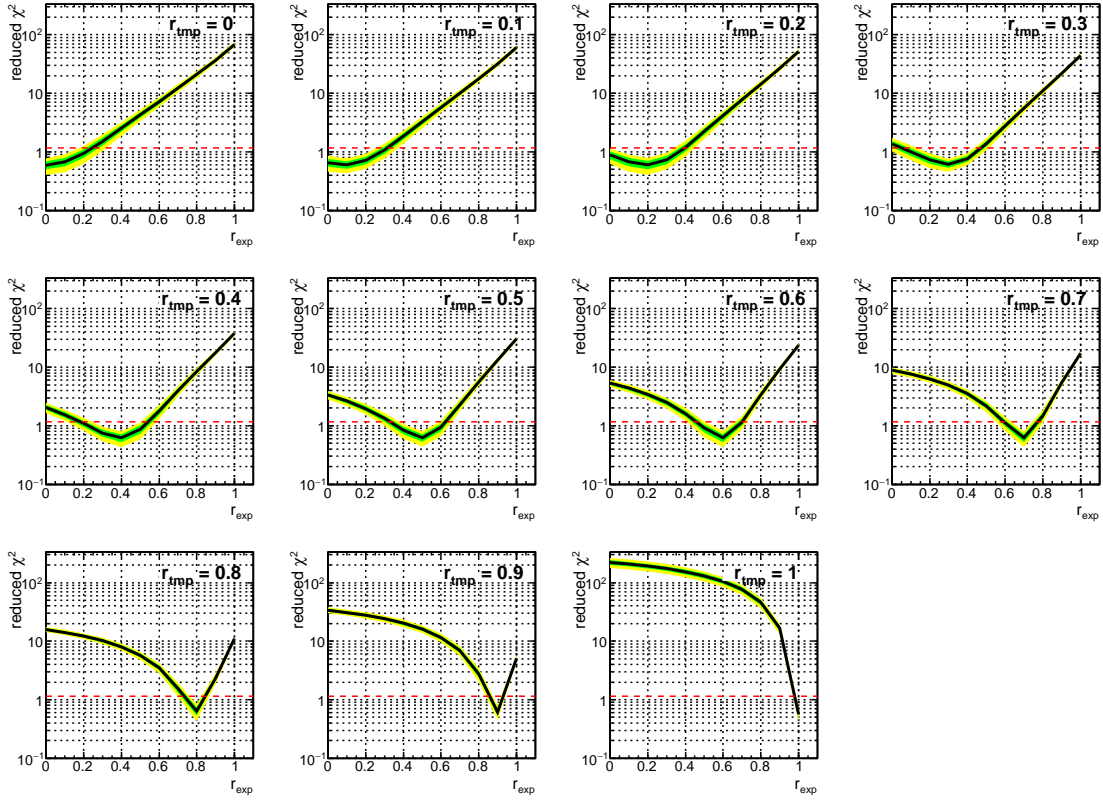} 
   \caption{Chi-squared test for target F. The pseudo-experimental data has $6\times10^3$ event number.}
   \label{fig:chi2F}
\end{figure}

\begin{figure}[htbp] %  figure placement: here, top, bottom, or page
   \centering
   \includegraphics[width=15cm]{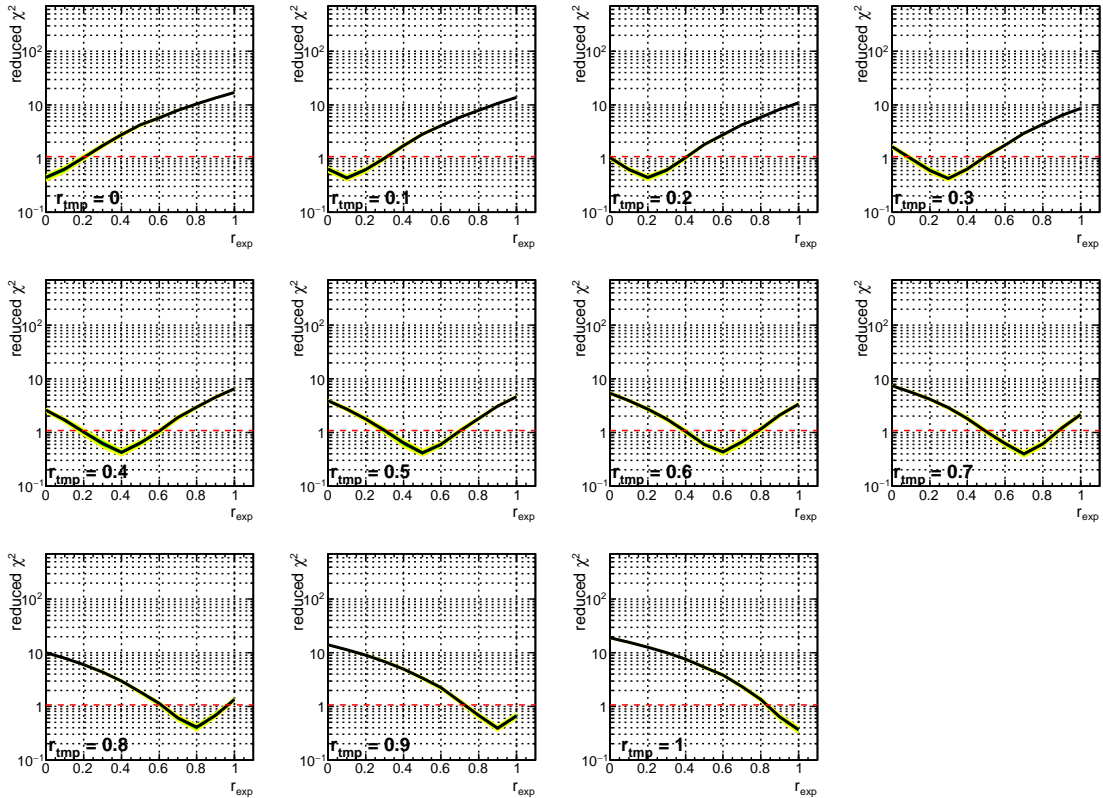} 
   \caption{Chi-squared test for target Ag. The pseudo-experimental data has $6\times10^4$ event number.}
   \label{fig:chi2Ag}
\end{figure}

In Figure \ref{fig:chi2F} and \ref{fig:chi2Ag}, results of chi-squared test between template data with particular $r$ and pseudo-experimental data
for F and Ag are shown, respectively. Red dashed line represents 90 \% confidence level (CL). 
In the figures, if anisotropic case suggested by N-body simulation $r=0.3$ is realized, completely isotropic case ($r=0$)
is rejected with $6\times10^3$ (for F) and $6\times10^4$ (for Ag) event numbers of the pseudo-experiment. 
The required event number depends on the energy threshold. Supposed energy thresholds are optimized to reduce the required 
event numbers of pseudo-experimental data.

\subsection{Angular histogram}
\label{subsec:histogram}
\begin{figure}[htbp] %  figure placement: here, top, bottom, or page
   \centering
   \includegraphics[width=15cm]{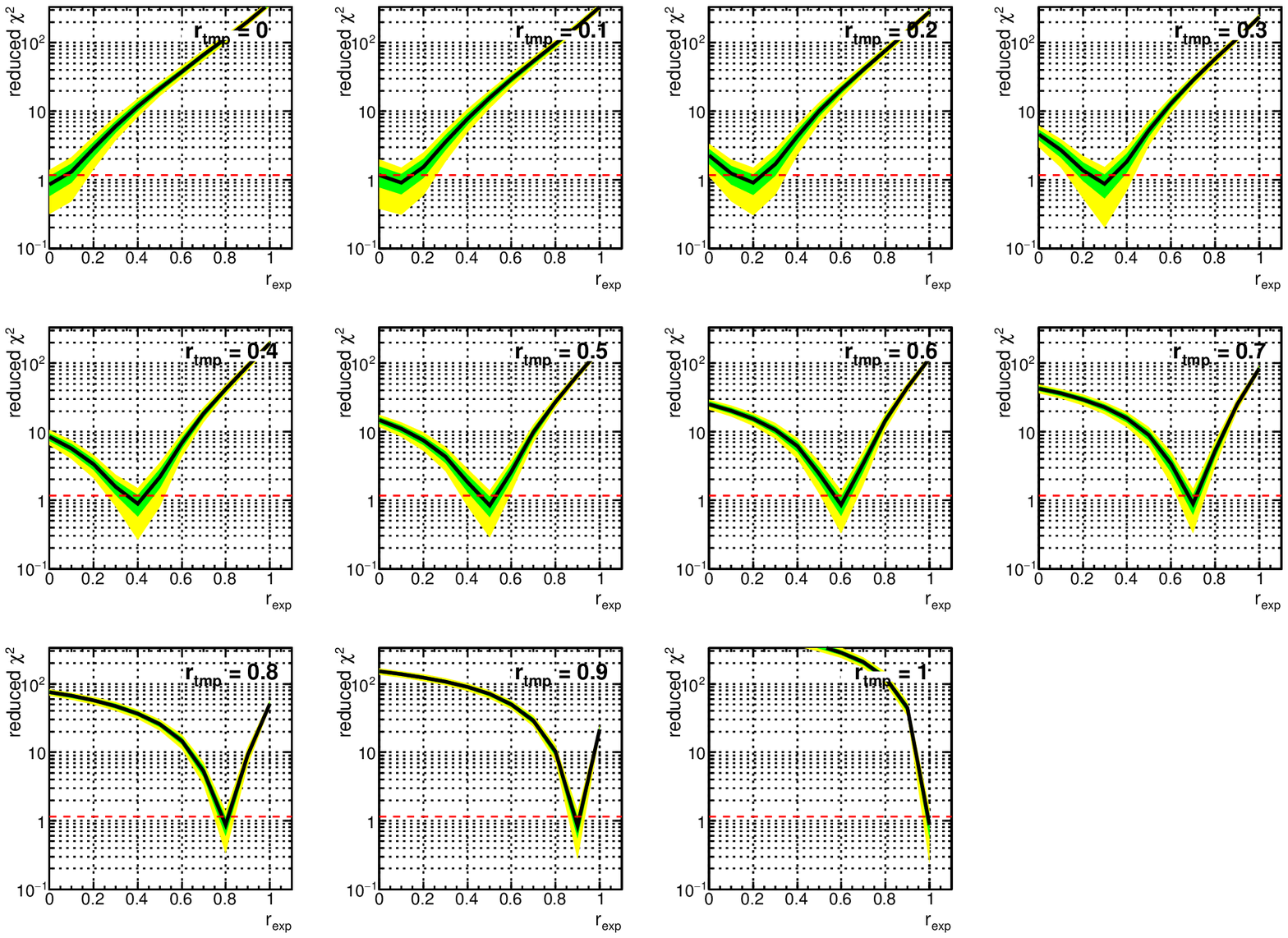} 
   \caption{Angular histogram for target F. The pseudo-experimental data has $5\times10^3$ event number.}
   \label{fig:angularhistF}
\end{figure}

\begin{figure}[htbp] %  figure placement: here, top, bottom, or page
   \centering
   \includegraphics[width=15cm]{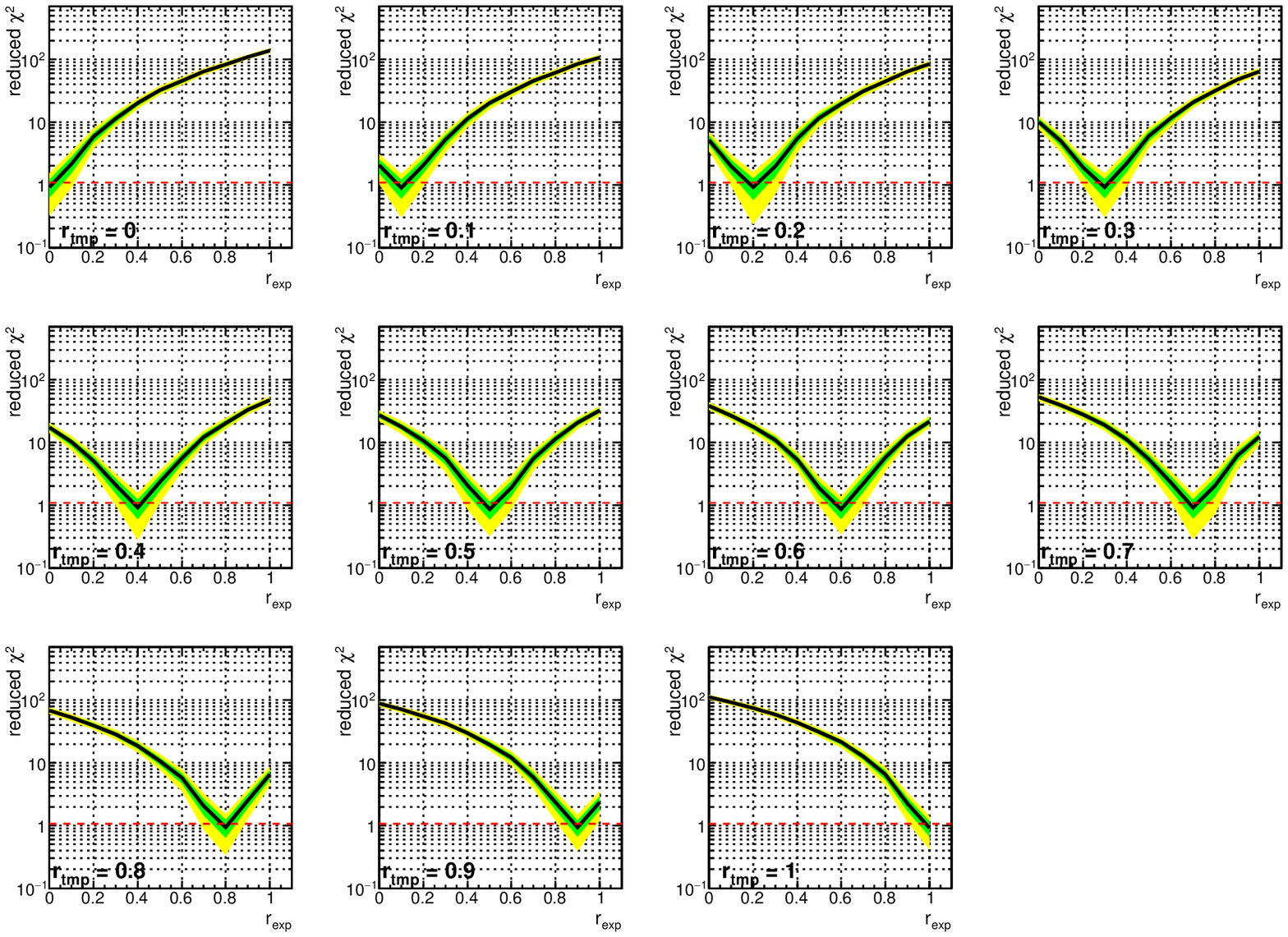} 
   \caption{Angular histogmra for target Ag. The pseudo-experimental data has $2\times10^4$ event number.}
   \label{fig:angularhistAg}
\end{figure}

Angular histogram is another candidate to analyze events in the directional detector. In Figure \ref{fig:angularhistF} and \ref{fig:angularhistAg}, 
chi-squared test of angular histogram for target F and Ag are shown, respectively. Red dashed line corresponds to at the 90\% CL. 
If anisotropic case ($r=0.3$) is realized, 
completely isotropic case ($r=0$) can be rejected at the 90\% CL with $5\times10^3$ (for F) and $2\times10^4$ (for Ag) event number. 
Since event number per a bin of energy-angular distribution is smaller than that of angular histogram, 
required event number is reduced compared to the energy-angular distribution.


\section{Conclusion}
\label{sec:conclusion}
Possibility to discriminate the anisotropy of dark matter velocity distribution using directional detector is investigated.
Depending on resolution of the detector, the energy-angular distribution and angular histogram can be analyzed. 
For energy-angular distribution case, $O(10^4)$ event number is required, while for angular histogram case $O(10^3-10^4)$ 
is required. It depends on the energy threshold of the detector which is determined by dark matter mass.

%%%%% acknowledgement %%%%%
%\vspace*{12pt}
%\noindent
%\\
%{\bf Acknowledgement}
%\vspace*{6pt}
%\noindent
%PPAP2018  is sponsored by JSPS Grant-in Aid for Scientific Research (C) Grant Number JP17K05418 and Core of Research for the Energetic Universe (Core-U) of Hiroshima University. 
%%%%% references %%%%%


\begin{thebibliography}{99}
%\cite{Nagao:2017yil}
\bibitem{Nagao:2017yil} 
  K.~I.~Nagao, R.~Yakabe, T.~Naka and K.~Miuchi,
  %``Discrimination of anisotropy in dark matter velocity distribution with directional detectors,''
  arXiv:1707.05523 [hep-ph].
  %%CITATION = ARXIV:1707.05523;%%
  %2 citations counted in INSPIRE as of 29 Apr 2018
\bibitem{darkdisk} 
J.~I.~Read, G.~Lake, O.~Agertz and V.~P.~Debattista,
  %``Thin, thick and dark discs in LCdark matter,''
  Mon.\ Not.\ Roy.\ Astron.\ Soc.\  {\bf 389}, 1041 (2008)
  doi:10.1111/j.1365-2966.2008.13643.x.
  %[arXiv:0803.2714 [astro-ph]].
  %%CITATION = doi:10.1111/j.1365-2966.2008.13643.x;%%
  %167 citations counted in INSPIRE as of 29 Aug 2016

\bibitem{darkdisk2} 
J.~I.~Read, L.~Mayer, A.~M.~Brooks, F.~Governato and G.~Lake,
  %``A dark matter disc in three cosmological simulations of Milky Way mass galaxies,''
  Mon.\ Not.\ Roy.\ Astron.\ Soc.\  {\bf 397}, 44 (2009)
  doi:10.1111/j.1365-2966.2009.14757.x.
 % [arXiv:0902.0009 [astro-ph.GA]].
  
\bibitem{LNAT} 
 F.~S.~Ling, E.~Nezri, E.~Athanassoula and R.~Teyssier,
  %``Dark Matter Direct Detection Signals inferred from a Cosmological N-body Simulation with Baryons,''
  JCAP {\bf 1002}, 012 (2010).
%  [arXiv:0909.2028 [astro-ph.GA]].
  %%CITATION = ARXIV:0909.2028;%%
%\cite{Kuhlen:2012fz}  
  
%\cite{Maciejewski:2010gz}
\bibitem{stream} 
M.~Maciejewski, M.~Vogelsberger, S.~D.~M.~White and V.~Springel,
  %``Bound and unbound substructures in Galaxy-scale Dark Matter haloes,''
  Mon.\ Not.\ Roy.\ Astron.\ Soc.\  {\bf 415}, 2475 (2011)
  doi:10.1111/j.1365-2966.2011.18871.x.
  %[arXiv:1010.2491 [astro-ph.CO]].
  %%CITATION = doi:10.1111/j.1365-2966.2011.18871.x;%%
  %23 citations counted in INSPIRE as of 29 Aug 2016

\bibitem{debris} 
M.~Lisanti and D.~N.~Spergel,
  %``Dark Matter Debris Flows in the Milky Way,''
  Phys.\ Dark Univ.\  {\bf 1}, 155 (2012)
  doi:10.1016/j.dark.2012.10.007.
 % [arXiv:1105.4166 [astro-ph.CO]].
  %%CITATION = doi:10.1016/j.dark.2012.10.007;%%
  %35 citations counted in INSPIRE as of 29 Aug 2016
  
\bibitem{KLS} 
 M.~Kuhlen, M.~Lisanti and D.~N.~Spergel,
  %``Direct Detection of Dark Matter Debris Flows,''
  Phys.\ Rev.\ D {\bf 86}, 063505 (2012)
  doi:10.1103/PhysRevD.86.063505.
  %[arXiv:1202.0007 [astro-ph.GA]].
  %%CITATION = doi:10.1103/PhysRevD.86.063505;%%
  %60 citations counted in INSPIRE as of 29 Aug 2016
\end{thebibliography}
\end{document}